\documentclass{svjour3}

\usepackage{amsmath,amssymb}
\usepackage{hyperref}
\usepackage{subfig}

\usepackage[amsmath, thmmarks]{ntheorem}
{\theoremstyle{nonumberplain}
\theoremsymbol{\ensuremath{\Box}}
\newtheorem{pf}{Proof}}

\newtheorem{thm}{Theorem}
\newtheorem{lem}{Lemma}
\newtheorem{cll}{Corollary}

\newtheorem{ex}{Example}

\title{A class of linear codes with few weights}

\author{Gaopeng Jian}    

\institute{Gaopeng Jian \at Key Laboratory of Machine Perception(MOE), School of EECS, Peking University, Beijing 100871, P.R.China; \\
\email{gpjian@pku.edu.cn}
}

\date{}

\begin{document}
%
%
%
  
  

\maketitle  

\begin{abstract}
Linear codes have been an interesting topic in both theory and practice for many years. In this paper, a class of $q$-ary linear codes with few weights are presented and their weight distributions are determined using Gauss periods.
Some of the linear codes obtained are optimal or almost optimal  with respect to the Griesmer bound.
As s applications, these linear codes can be used to construct secret sharing schemes with nice access structures.
\keywords{Linear codes \and Weight distribution \and Gauss periods}
\end{abstract}



\section{Introduction}
Let $q$ be a power of a prime and $\mathbb{F}_q$ denote the finite field with $q$ elements.
An $[n,k,d]$ linear code $C$ over $\mathbb{F}_q$ is a $k$-dimensional subspace of $\mathbb{F}_q^n$ with minimum Hamming distance $d$.
Let $A_i$ be the number of codewords with Hamming weight $i$ in $C$. 
The polynomial $1+A_1 z+A_2 z^2+\cdots+A_n z^n$ is called the weight enumerator of $C$ and the sequence $(1, A_1, A_2, \ldots, A_n)$ called the weight distribution of $C$.
If the number of nonzero $A_i$ in the sequence $(A_1, A_2, \ldots, A_n)$ is equal to $t$, we call $C$ a $t$-weight code.
The weight distribution of a code contains important information on its error correcting capability and the error probability of its error detection and correction with respect to some algorithms \cite{torleiv2007codes}.
In addition, much attention has been paid to two-weight and three-weight linear codes  \cite{tang2016linear,zhou2015linear,heng2016three,heng2016two,heng2015class,heng2017construction,luo2018binary,ding2014binary,ding2015class} due to their applications in secret sharing schemes \cite{yuan2006secret,carlet2005linear}, strongly regular graphs \cite{calderbank1986geometry}, association schemes \cite{calderbank1984three} and authentication codes \cite{ding2005coding}.  

We recall the Griesmer bound for linear codes in the following lemma \cite{griesmer1960bound}.
\begin{lem}[Griesmer bound]  
For any $[n,k,d]$ code over $\mathbb{F}_q$, we have
\[
n \ge \sum_{i=0}^{k-1} \left \lceil \frac{d}{q^i} \right \rceil,
\]
where $\lceil \cdot \rceil$  is the ceiling function.
\end{lem}

Let $\text{Tr}^m_1(\cdot)$ denote the trace function from $\mathbb{F}_{q^m}$ to $\mathbb{F}_q$, and let $D=\{ d_1, d_2, \ldots, d_n \}$ be a subset of $\mathbb{F}_{q^m}^*$. 
Define a linear code of length $n = |D|$ over $\mathbb{F}_q$ by
\begin{equation} \label{d1}
C_D=\{(\text{Tr}^m_1(ad_1), \text{Tr}^m_1(ad_2), \ldots, \text{Tr}^m_1(ad_n)): a \in \mathbb{F}_{q^m} \}.
\end{equation}
The set $D$ is called the \emph{defining set} of the trace code $C_D$.
The construction was first proposed by Ding et al. in  \cite{ding2007cyclotomic} and many classes of known codes could be produced by properly selecting the defining set.
Recently, researchers have extended the construction to codes over finite rings and obtain many classes of linear codes with few weights by the Gray map \cite{luo2018five,liu2017several,shi2016optimal,shi2017two,shi2016new,liu2018two}.

Let $m_1$ be a positive divisor of $m$, and let $\text{Tr}^{m_1}_1(\cdot)$ denote the trace function from $\mathbb{F}_{q^{m_1}}$ to $\mathbb{F}_q$.
For $K \subset \mathbb{F}_{q^{m_1}} \times \mathbb{F}_{q^m}^*$, Liu et al. \cite{liu2018several} defined a linear code of length $2|K|$ over $\mathbb{F}_q$ by
\begin{equation} \label{e1}
C_K=\{c(a,b): a \in \mathbb{F}_{q^{m_1}}, b \in \mathbb{F}_{q^m} \}, 
\end{equation}
where
\[
c(a,b)=(\text{Tr}^{m_1}_1(ax)+\text{Tr}^m_1(by)|\text{Tr}^{m_1}_1(ax)+\text{Tr}^m_1((a+b)y))_{(x,y) \in K}.
\]
Note that the Gray image of trace codes over the ring $R=\mathbb{F}_q+u\mathbb{F}_q$ with $u^2=0$ coincides with the construction when $m=m_1$.

Let $\omega$ be a fixed primitive element of $\mathbb{F}_{q^m}$, and let $D=\{1,\omega^h,\ldots,\omega^{(n-1)h} \}$, where $h | \frac{q^m-1}{q-1}$ and $\frac{q^m-1}{h(q-1)} |n$.
Note that if $n=\frac{q^m-1}{h}$, then $D=C_0^{(h,q^m)}$ and the linear code $C_D$ defined by \eqref{d1} is equivalent to an irreducible cyclic codes over $\mathbb{F}_q$, which is not true in general \cite{ding2008two,heng2018construction}.
In this paper we will investigate the weight distribution of $C_K$ defined by \eqref{e1} with $K=\mathbb{F}_{q^{m_1}} \times D$.
Several classes of two-weight and three-weight linear codes are derived employing Gauss periods over finite fields.
In particular, the two-weight codes are optimal with respect to the Griesmer bound if $n$ is small.
Furthermore, an application to secret sharing schemes is sketched out.

This paper is organized as follows. 
In Section \ref{pl}, we briefly recall some definitions and notations which will be used later.  
In Section \ref{wt}, we investigate the weight distribution of $C_K$ by using Gauss periods and construct two-weight and three-weight codes over $\mathbb{F}_q$. 
In Section \ref{sss}, we state an application in secret sharing schemes.
In Section \ref{clu}, we conclude this paper.

\section{Preliminaries} \label{pl}
In this section,  we recall some basic results of characters, cyclotomic classes and Gauss periods. Interested readers are referred to \cite{lidl1997finite,myerson1981period} for more details.

Suppose that $q=p^s$ for an odd prime $p$ and a positive integer $s$. For $a \in \mathbb{F}_q$, an additive character of the finite field $\mathbb{F}_q$ can be defined by
\[
\chi_a: \mathbb{F}_q \rightarrow \mathbb{C}^*, \chi_a(x)=\zeta_p^{\text{Tr}(ax)},
\]
where $\zeta_p=e^{\frac{2 \pi \sqrt{-1}}{p}}$ is a primitive $p$-th root of unity and $\text{Tr}(\cdot)$ denotes the trace function from $\mathbb{F}_q$ onto $\mathbb{F}_p$. 
It is clear that $\chi_0(x) = 1$ for all $x \in \mathbb{F}_q$ and $\chi_0$ is called the \emph{trivial additive character} of $\mathbb{F}_q$. If $a=1$, we call $\chi:=\chi_1$ the \emph{canonical additive character} of $\mathbb{F}_q$. 
The orthogonal property of additive characters is given as
\[
\sum_{x \in \mathbb{F}_q} \chi(ax)=\begin{cases}
q, & \text{if } a=0, \\
0, & \text{otherwise}.
\end{cases}
\] 

Let $\alpha$ be a fixed primitive element of $\mathbb{F}_q$. For a positive integer $N>1$ such that $N|(q-1)$, we define
\[
C_i^{(N,q)}=\alpha^i \langle \alpha^N \rangle, \ i=0,1,\ldots,N-1,
\]
where $\langle \alpha^N \rangle$ denotes the cyclic subgroup of $\mathbb{F}_q^*$ generated by $\alpha^N$.
The cosets $C_i^{(N,q)}$ are called the \emph{cyclotomic classes} of order $N$ in  $\mathbb{F}_q$. 

The \emph{Gaussian periods} are defined by
\[
\eta_i^{(N,q)}=\sum_{x \in C_i^{(N,q)}} \chi(x),\ i=0,1,\ldots,N-1,
\]
where $\chi$ is the canonical additive character of $\mathbb{F}_q$.

\begin{lem} \label{l3}
\[
\left | \eta_i^{(N,q)}+\frac{1}{N} \right | \le \frac{(N-1)\sqrt{q}}{N}. 
\]
\end{lem}

\begin{lem}  \label{l2}
Let $N=2$, then the Gaussian periods are given by
\[
\eta_0^{(2,q)}=\begin{cases}
\frac{-1+(-1)^{s-1}\sqrt{q}}{2}, & \text{if} \ p \equiv 1 \pmod{4},\\
\frac{-1+(-1)^{s-1}(\sqrt{-1})^{s} \sqrt{q}}{2}, & \text{if} \ p \equiv 3 \pmod{4}
\end{cases}
\]
and
\[
\eta_1^{(2,q)}=-1-\eta_0^{(2,q)}.
\]
\end{lem}

\begin{lem}[The semiprimitive case] \label{l7}
Assume that $N \ge 2$, $q=p^{2j\gamma}$, where $N | (p^j+1)$ and $j$ is the smallest such positive integer. Then the Gaussian periods of order $N$ are given below:
\begin{description}
\item[(a)] If $\gamma, p, \frac{p^j+1}{N}$ are all odd, then
\[
\eta_{\frac{N}{2}}^{(N,q)}=\sqrt{q}-\frac{\sqrt{q}+1}{N}, \
\eta_i^{(N,q)}=-\frac{\sqrt{q}+1}{N}  \text{ for all } i \neq \frac{N}{2}.
\]
\item[(b)] In all the other cases,
\[
\eta_0^{(N,q)}=\frac{(-1)^{\gamma+1}(N-1)\sqrt{q}-1}{N}, \
\eta_i^{(N,q)}=\frac{(-1)^{\gamma}\sqrt{q}-1}{N} \text{ for all } i \neq 0.
\]
\end{description}
\end{lem}

\section{The weight distribution of  $C_K$} \label{wt}
In this section, we investigate the weight distribution of $C_K$ defined by \eqref{e1}.
Let $\mathbb{F}_{q^m}^*=\langle \omega \rangle$ and $D=\{\omega^{hi}:0 \le i<n \}$ where $h | \frac{q^m-1}{q-1}$ and $\frac{q^m-1}{h(q-1)}|n$.
We know that any codeword of $C_K$ has the following form:
\[
c(a,b)=(\text{Tr}^{m_1}_1(ax)+\text{Tr}^m_1(by)|\text{Tr}^{m_1}_1(ax)+\text{Tr}^m_1((a+b)y))_{(x,y) \in K},
\]
where $K=\mathbb{F}_{q^{m_1}} \times D$ and $a \in \mathbb{F}_{q^{m_1}}$, $b \in \mathbb{F}_{q^m}$.
The Hamming weight of $c(a,b)$ is expressed in terms of  Gauss periods below.
\begin{thm} \label{th}
For $a \in \mathbb{F}_{q^{m_1}}$, $b \in \mathbb{F}_{q^m}$, the Hamming weight $w_H(c(a,b))$ is 
\begin{enumerate}
\item If $a=b=0$, then $w_H(c(a,b))=0$;
\item If $a=0$ and $b \in C_i^{(h,q^m)}$, $i=0,1,\ldots,h-1$, then 
\[
w_H(c(a,b))=\frac{2nq^{m_1-1}(q-1)}{q^m-1} \left( q^m-1-h\eta_i^{(h,q^m)} \right);
\]
\item If $a \in \mathbb{F}_{q^{m_1}}^*$ and $b \in \mathbb{F}_{q^m}$, then $w_H(c(a,b))=2nq^{m_1-1}(q-1)$.
\end{enumerate}
\end{thm}
\begin{pf}
Let $\chi$, $\chi^{(m)}$, $\chi^{(m_1)}$ be the canonical additive characters of $\mathbb{F}_q$, $\mathbb{F}_{q^m}$ and $\mathbb{F}_{q^{m_1}}$ respectively.
Let $N$ be a positive integer. For any vector $\mathbf{y}=(y_1,y_2,\ldots,y_N) \in \mathbb{F}_q^N$, let
\[
\Psi(\mathbf{y})=\sum_{i=1}^N \chi(y_i).
\]
By the orthogonal property of additive characters
\begin{align}
w_H(\mathbf{y}) &=N-\frac{1}{q}\sum_{i=1}^N \sum_{s \in \mathbb{F}_q} \chi(sy_i) \notag \\
&=(1-\frac{1}{q})N-\frac{1}{q} \sum_{s \in \mathbb{F}_q^*} \Psi(s\mathbf{y}) \label{wh}
\end{align}
If $a \in \mathbb{F}_{q^{m_1}}^*$ and $b \in \mathbb{F}_{q^m}$,
\begin{align}
\Psi(c(a,b))= &\sum_{x \in \mathbb{F}_{q^{m_1}}} \sum_{y \in D} \chi(\text{Tr}^{m_1}_1(ax)+\text{Tr}^m_1(by)) \notag \\
&+\sum_{x \in \mathbb{F}_{q^{m_1}}} \sum_{y \in D} \chi(\text{Tr}^{m_1}_1(ax)+\text{Tr}^m_1((a+b)y)) \notag\\
=&\sum_{y \in D}\chi^{(m)}(by)\sum_{x \in \mathbb{F}_{q^{m_1}}}\chi^{(m_1)}(ax) \notag\\
&+\sum_{y \in D}\chi^{(m)}((a+b)y)\sum_{x \in \mathbb{F}_{q^{m_1}}}\chi^{(m_1)}(ax) \notag \\
=&0. \label{w2}
\end{align}
If $a=0$, then $c(0,b) = (\text{Tr}^m_1(by)|\text{Tr}^m_1(by))_{(x,y) \in K}$ and
\begin{align*}
\sum_{s \in \mathbb{F}_q^*}\Psi(s \cdot c(0,b)) &=2\sum_{s \in \mathbb{F}_q^*}\sum_{x \in \mathbb{F}_{q^{m_1}}} \sum_{y \in D} \chi(s\text{Tr}^m_1(by)) \\
&=2q^{m_1}\sum_{s \in \mathbb{F}_q^*}\sum_{i=0}^{n-1}\chi^{(m)}(sb\omega^{hi}) 
\end{align*}
Since $h | \frac{q^m-1}{q-1}$ and $\frac{q^m-1}{h(q-1)}|n$, then 
\[
C_0^{(h,q^m)}=\langle \omega^h \rangle=\bigcup_{j=0}^{\frac{q^m-1}{h(q-1)}-1} \omega^{hj}\langle \omega^{\frac{q^m-1}{q-1}} \rangle=\bigcup_{j=0}^{\frac{q^m-1}{h(q-1)}-1} \omega^{hj} \mathbb{F}_q^*.
\]
and for $b \in C_i^{(h,q^m)}$,
\begin{align}
\sum_{s \in \mathbb{F}_q^*}\Psi(s \cdot c(0,b)) &=2q^{m_1} \sum_{s \in \mathbb{F}_q^*} \sum_{l=0}^{\frac{hn(q-1)}{q^m-1}-1} \sum_{j=0}^{\frac{q^m-1}{h(q-1)}-1} \chi^{(m)}(sb\omega^{hj+\frac{q^m-1}{q-1}l}) \notag \\
&=2q^{m_1} \sum_{l=0}^{\frac{hn(q-1)}{q^m-1}-1} \sum_{j=0}^{\frac{q^m-1}{h(q-1)}-1} \sum_{s \in \mathbb{F}_q^*} \chi^{(m)}((s\omega^{hj})b\omega^{\frac{q^m-1}{q-1}l}) \notag \\
&=2q^{m_1} \sum_{l=0}^{\frac{hn(q-1)}{q^m-1}-1} \sum_{t \in C_0^{(h,q^m)}} \chi^{(m)}(bt(\omega^h)^{\frac{q^m-1}{h(q-1)}l}) \notag \\
&=\frac{2q^{m_1}hn(q-1)}{q^m-1} \sum_{t \in C_0^{(h,q^m)}} \chi^{(m)}(bt) \notag \\
&=\frac{2q^{m_1}hn(q-1)}{q^m-1}\cdot \eta_i^{(h,q^m)}. \label{w1}
\end{align}
By \eqref{wh},\eqref{w2} and \eqref{w1} we complete the proof.
\end{pf}

\begin{cll}
If $h<q^{m/2}+1$, $C_K$ has  parameters
\[
\left[ 2nq^{m_1},m+m_1,d \ge \left \lceil \frac{2nq^{m/2+m_1-1}(q-1)}{q^m-1} (q^{m/2}+1-h) \right \rceil \right].
\]
\end{cll}
\begin{pf}
By Theorem \ref{th} and Lemma \ref{l3}, for $(a,b) \ne (0,0)$,
\[
w_H(c(a,b)) \ge \left \lceil \frac{2nq^{m_1-1}(q-1)}{q^m-1} (q^m-(h-1)q^{m/2}) \right \rceil>0
\]
as $h<q^{m/2}+1$.
Thus the dimension of $C_K$ is $m+m_1$.
\end{pf}

\begin{cll}\label{h1}
If $h=1$, $C_K$ is a $[2nq^{m_1},m+m_1,2nq^{m_1-1}(q-1)]$ two-weight linear code with the weight distribution given in Table \ref{t1}.
Furthermore, $C_K$ meets the Griesmer bound if $n<\frac{q(q^m-1)}{2(q-1)}$.
\end{cll}
\begin{pf}
If $h=1$, for $a=0$ and $b \in \mathbb{F}_{q^m}^*$,
\[
w_H(c(a,b))=\frac{2nq^{m_1-1}(q-1)}{q^m-1} \left( q^m-1-\eta_0^{(1,q^m)} \right)=\frac{2nq^{m_1+m-1}(q-1)}{q^m-1}.
\]
We can obtain the parameters and weight distribution of $C_K$ from Theorem \ref{th}.

If $n<\frac{q(q^m-1)}{2(q-1)}$, let $n=\frac{t(q^m-1)}{q-1}$ with $0<t<\frac{q}{2}$, then
\begin{align*}
\sum_{i=0}^{m+m_1-1} \left \lceil \frac{2nq^{m_1-1}(q-1)}{q^i} \right \rceil=& 2n(q-1)\sum_{i=0}^{m_1-1}q^{m_1-1-i}+\sum_{i=m_1}^{m+m_1-1} \left \lceil \frac{2t(q^m-1)}{q^{i+1-m_1}} \right \rceil \\
=& 2n(q^{m_1}-1)+\sum_{j=1}^m \left \lceil 2tq^{m-j}-\frac{2t}{q^j} \right \rceil \\
=&2n(q^{m_1}-1)+2t\sum_{j=1}^m q^{m-j} \\
=& 2nq^{m_1}.
\end{align*}
Thus $C_K$ meets the Griesmer bound.
\end{pf}

\begin{table}
\caption{Weight distribution for $h=1$} 
\centering
\begin{tabular}{ll} 
\hline
Weight                & Frequency \\
\hline
0 &1 \\
$2nq^{m_1-1}(q-1)$ & $q^{m_1+m}-q^m$\\
$\frac{2nq^{m_1+m-1}(q-1)}{q^m-1}$ & $q^m-1$\\
\hline
\end{tabular}
\label{t1}
\end{table}

\begin{ex}
Let $(q,m,m_1,h,n)=(3,2,1,1,4)$.
Then $C_K$ has parameters [24,3,16] and weight enumerator $1 + 18z^{16} + 8z^{18}$.
The code is optimal.
As a comparison, it is different from the best known linear codes from the Magma BKLC(GF(3),24,3) which has a different weight enumerator $1 + 12z^{16} + 12z^{17}+2z^{18}$.
\end{ex}

\begin{ex}
Let $(q,m,m_1,h,n)=(3,2,1,1,8)$.
Then $C_K$ has parameters [48,3,32] and weight enumerator $1 + 18z^{32} + 8z^{36}$.
The code is almost optimal as the best linear code of length 48 and dimension 3 over $\mathbb{F}_3$ has minimum weight 33.
\end{ex}

\begin{ex}
Let $(q,m,m_1,h,n)=(3,2,2,1,4)$.
Then $C_K$ has parameters [72,4,48] and weight enumerator $1 + 72z^{48} + 8z^{54}$.
The code is optimal.
As a comparison, it is different from the best known linear codes from the Magma BKLC(GF(3),72,4) which has a different weight enumerator $1 + 66z^{48} + 12z^{51}+2z^{54}$.
\end{ex}

\begin{cll}\label{h2}
If $h=2$, $C_K$ is a $[2nq^{m_1},m+m_1,\frac{2nq^{m/2+m_1-1}(q-1)}{q^m-1}(q^{m/2}-1)]$ three-weight linear code with the weight distribution given in Table \ref{t2}.
\end{cll}
\begin{pf}
Since $2|\frac{q^m-1}{q-1}$, $m$ is even.
For $a=0$ and $b \in C_i^{(2,q^m)}$, $i=0,1$, 
\[
w_H(c(a,b))=\frac{2nq^{m_1-1}(q-1)}{q^m-1}\left(q^m-(-1)^i(1+2\eta_0^{(2,q^m)})\right)
\]
and $|2\eta_0^{(2,q^m)}+1|=q^{m/2}$ by Lemma \ref{l2}.
We can obtain the parameters and weight distribution of $C_K$ from Theorem \ref{th}.
\end{pf}

\begin{table}
\caption{Weight distribution for $h=2$} 
\centering
\begin{tabular}{ll} 
\hline
Weight                & Frequency \\
\hline
0 &1 \\
$2nq^{m_1-1}(q-1)$ & $q^{m_1+m}-q^m$\\
$\frac{2nq^{m/2+m_1-1}(q-1)}{q^m-1}(q^{m/2}-1)$ & $\frac{q^m-1}{2}$\\
$\frac{2nq^{m/2+m_1-1}(q-1)}{q^m-1}(q^{m/2}+1)$ & $\frac{q^m-1}{2}$\\
\hline
\end{tabular}
\label{t2}
\end{table}

\begin{ex}
Let $(q,m,m_1,h,n)=(3,2,1,2,4)$.
Then $C_K$ has parameters [24,3,12] and weight enumerator $1 + 4z^{12} + 18z^{16} + 4z^{24}$.
\end{ex}

\begin{ex}
Let $(q,m,m_1,h,n)=(3,2,2,2,4)$.
Then $C_K$ has parameters [72,4,36] and weight enumerator $1 + 4z^{36} + 72z^{48} + 4z^{72}$.
\end{ex}

\begin{cll}\label{h3}
Assume that $2 \le h<q^{m/2}+1$, $q^m= p^{2j \gamma}$, where $h|(p^j +1)$ and $j$ is the smallest such positive integer. Then $C_K$ is a $[2nq^{m_1},m+m_1,d]$ three-weight linear code with the weight distribution given in Table \ref{t3}, where 
\[
d=\begin{cases} 
\frac{2nq^{m/2+m_1-1}(q-1)}{q^m-1}(q^{m/2}-1) & \text{if $\gamma$ is even} \\
\frac{2nq^{m/2+m_1-1}(q-1)}{q^m-1}(q^{m/2}-h+1) & \text{if $\gamma$ is odd.}
\end{cases}
\]
\end{cll}
\begin{pf}
By Lemma \ref{l7}, $\{\eta_i^{(h,q^m)}:i=0,1,\ldots,h-1\}$ takes value $\frac{(-1)^{\gamma+1}(h-1)q^{m/2}-1}{h}$ with multiplicity 1 and $\frac{(-1)^{\gamma}q^{m/2}-1}{h}$ with multiplicity $h-1$.
We can obtain the parameters and weight distribution of $C_K$ from Theorem \ref{th}.
\end{pf}

\begin{table}
\caption{Weight distribution in the semiprimitive case} 
\centering
\begin{tabular}{ll} 
\hline
Weight                & Frequency \\
\hline
0 &1 \\
$2nq^{m_1-1}(q-1)$ & $q^{m_1+m}-q^m$\\
$\frac{2nq^{m/2+m_1-1}(q-1)}{q^m-1}(q^{m/2}+(-1)^{\gamma}(h-1))$ & $\frac{q^m-1}{h}$\\
$\frac{2nq^{m/2+m_1-1}(q-1)}{q^m-1}(q^{m/2}+(-1)^{\gamma+1})$ & $\frac{(q^m-1)(h-1)}{h}$\\
\hline
\end{tabular}
\label{t3}
\end{table}

\begin{ex}
Let $(q,m,m_1,h,n)=(3,4,2,4,10)$. Then $j=1$ and $\gamma=2$. 
The code $C_K$ has parameters [180,6,108] and weight enumerator $1 + 60z^{108} + 648z^{120} + 20z^{162}$.
\end{ex}

\begin{ex}
Let $(q,m,m_1,h,n)=(3,4,2,5,8)$. Then $j=2$ and $\gamma=1$.
The code $C_K$ has parameters [144,6,54] and weight enumerator $1 + 16z^{54} + 648z^{96} + 64z^{108}$.
\end{ex}

\section{Applications in secret sharing schemes} \label{sss}
For a vector $\mathbf{x} \in \mathbb{F}_q^n$, the support $s(\mathbf{x})$ of $\mathbf{x}$ is defined as the set of indices where it is nonzero.
We say that a vector $\mathbf{x}$ covers a vector $\mathbf{y}$ if $s(\mathbf{x})$ contains $s(\mathbf{y})$.
For a linear code $C$ over $\mathbb{F}_q$, a codeword $\mathbf{c} \in C$ is minimal if it covers only codewords of the form $a \cdot \mathbf{c}$, where $a \in \mathbb{F}_q$.
$C$ is minimal if every codeword of $C$ is minimal.
If the weights of $C$ are close enough to each other, then $C$ is minimal, as described by the following lemma \cite{ashikhmin1998minimal}.

\begin{lem}
Denote by $w_{min}$ and $w_{max}$ the minimum and maximum nonzero weight of a given $q$-ary linear code $C$, respectively. If $w_{min}/w_{max}>\frac{q-1}{q}$, then $C$ is minimal.
\end{lem}

The notion of minimal codewords was introduced to determine the set of all minimal access sets of a secret sharing scheme (SSS).
Massey's scheme is a construction of a SSS based on coding theory \cite{yuan2006secret}.
When  $C$ is minimal, it was stated in \cite{ding2003covering} that there is the following alternative, depending on $d^{\perp}$ (the minimum distance of the dual of $C$):
\begin{itemize}
\item If $d^{\perp} \ge 3$, then the SSS is \emph{``democratic''}: every user belongs to the same number of coalitions,
\item If $d^{\perp}=2$, then there are users who belong to every coalition: the \emph{``dictators''}.
\end{itemize}
It's easy to check the following:
\begin{enumerate}
\item The code $C_K$ in Corollary \ref{h1} is minimal, provided that $m > 1$. 
\item The code $C_K$ in Corollary \ref{h2} is minimal, provided that $m > 2$.
\item The code $C_K$ in Corollary \ref{h3} is minimal, provided that $\gamma$ is even and $h(q-1)<q^{m/2}-1$ or $\gamma$ is odd and $hq <q^{m/2}+1$.
\end{enumerate}
Besides, a SSS built on $C_K$ is dictatorial as the minimum distance of the dual of $C_K$ is 2, which can be seen from the nondegenerate property of the trace function and the observation below: 

Since $\omega^{\frac{q^m-1}{q-1}} \in D$, there exist $(x_1,y_1),(x_2,y_2) \in K$ such that $x_1/x_2=y_1/y_2=\beta \in \mathbb{F}_q^*$. Then $\text{Tr}^{m_1}_1(ax_1)+\text{Tr}^m_1(by_1)=\beta \left(\text{Tr}^{m_1}_1(ax_2)+\text{Tr}^m_1(by_2) \right)$ for all $a \in \mathbb{F}_{q^{m_1}}$, $b \in \mathbb{F}_{q^m}$.

\section{Concluding remarks} \label{clu}
In this paper, inspired by the work in \cite{liu2017several} and \cite{liu2018several}, we presented a class of linear codes over $\mathbb{F}_q$ and determined their weight distributions using Gauss periods.
Our results showed that the presented linear codes have few weights and some of them are optimal or almost optimal with respect to the Griesmer bound.
It would be interesting if more linear codes with few weights can be presented.


\bibliographystyle{plain} 
\bibliography{thesis} 
\end{document}